\documentclass{article}

     \PassOptionsToPackage{numbers, compress}{natbib}

\usepackage[final]{neurips_2024}

\usepackage[utf8]{inputenc} 
\usepackage[T1]{fontenc}    
\usepackage{hyperref}       
\usepackage{url}            
\usepackage{booktabs}       
\usepackage{amsfonts}       
\usepackage{nicefrac}       
\usepackage{microtype}      
\usepackage{xcolor}         
\usepackage{longtable}
\usepackage{pdflscape}
\usepackage{array}
\usepackage{tcolorbox}
\usepackage{enumitem}
\usepackage{amsmath}
\usepackage{amssymb}
\usepackage{graphicx}
\usepackage{hyperref}
\usepackage{tcolorbox}
\usepackage{amsmath}
\usepackage{mdframed}
\usepackage[utf8]{inputenc} 
\usepackage[T1]{fontenc}    
\usepackage{hyperref}       
\usepackage{url}            
\usepackage{booktabs}       
\usepackage{amsfonts}       
\usepackage{nicefrac}       
\usepackage{microtype}      
\usepackage{xcolor}         
\usepackage{tcolorbox}
\tcbuselibrary{skins}

\title{GPAI Evaluations Standards Taskforce: Towards Effective AI Governance  \thanks{The authors thank Anthony Barrett, Ben Bucknall, Herbie Bradley, Simeon Campos, Stephen Casper, Connor Dunlop, Jimmy Farrell, Arthur Goemans, Koen Holtman, David Manheim, Lorenzo Pacchiardi, Gabriel Sander, and Gaurav Sett for feedback and comments; and the Technology and Security Policy (TASP) leadership team, especially Michael Aird, Jeff Alstott, Chris Byrd, Casey Dugan, Ella Guest, and Brodi Kotila for their support.} \thanks{Funding for this work was provided by gifts from RAND supporters.} \thanks{RAND working papers are intended to share early insights and solicit informal peer review. This working paper has been approved for circulation by RAND Global and Emerging Risks but has not yet completed formal peer review or been professionally edited or proofread. This working paper can be quoted and cited without permission of the author, provided the source is clearly referred to as a working paper. This working paper does not necessarily reflect the opinions of RAND’s research clients and sponsors.}}

\author{%
  Patricia Paskov, RAND \\
  \texttt{ppaskov@rand.org} \\
    \And
  Lukas Berglund, RAND \\
  \texttt{lberglund@rand.org} \\  
  \And
  Everett Smith, RAND \\
  \texttt{everetts@rand.org}\thanks{Patricia Paskov, Lukas Berglund, and Everett Smith are Technology and Security Policy (TASP) fellows at RAND; for more information on the fellowship program, visit www.rand.org/tasp-fellows.} \\  
    \And
  Lisa Soder, Interface \\
  \texttt{lsoder@interface-eu.org} \\  
   \\
}
\begin{document}
\maketitle
\begin{abstract}
General-purpose AI (GPAI) evaluations have been proposed as a promising way of identifying and mitigating systemic risks posed by AI development and deployment. While GPAI evaluations play an increasingly central role in institutional decision- and policy-making – including by way of the European Union (EU) AI Act’s mandate to conduct evaluations on GPAI models presenting systemic risk – no standards exist to date to promote their quality or legitimacy. To strengthen GPAI evaluations in the EU, which currently constitutes the first and only jurisdiction that mandates GPAI evaluations, we outline four desiderata for GPAI evaluations: internal validity, external validity, reproducibility, and portability. To uphold these desiderata in a dynamic environment of continuously evolving risks, we propose a dedicated EU GPAI Evaluation Standards Taskforce, to be housed within the bodies established by the EU AI Act. We outline the responsibilities of the Taskforce, specify the GPAI provider commitments that would facilitate Taskforce success, discuss the potential impact of the Taskforce on global AI governance, and address potential sources of failure that policymakers should heed.
\end{abstract}

\section{Introduction}
\textbf{General-purpose AI (GPAI) evaluations have emerged as a valuable tool for measuring and addressing systemic risks posed by AI development and deployment.} While GPAI evaluations play an increasingly central role in institutional decision- and policy-making – including by way of the European Union (EU) AI Act’s mandate to conduct evaluations on GPAI models presenting systemic risk \cite{EUAIAct2024}– no standards exist to date to promote their quality or legitimacy. This paper puts forth desiderata for GPAI evaluations and proposes a dedicated EU GPAI Evaluation Standards Taskforce to promote these desiderata adaptably in a dynamic environment of continuously evolving risks. This paper focuses specifically on an EU Taskforce, given the EU is, at the time of writing, the only jurisdiction with the power to mandate and regulate GPAI evaluation practices. A Taskforce in the EU, therefore, warrants timely consideration and enables direct policy influence. Actions in the EU now could plausibly pave the pathway for coordinated international standards in the future.  

\textbf{This paper is relevant to parties involved in or affected by decision-making processes surrounding GPAI evaluations, particularly in the EU. } Primary audiences include the EU AI Office and the Codes of Practice working group chairs and participants. The paper’s implications, however, reach beyond the EU, and towards international standards more broadly. Additional audiences may therefore include AI Safety Institutes (AISIs), civil society groups focused on AI governance, GPAI evaluation organisations, GPAI providers, independent researchers, and international standards organisations. 

\textbf{The paper proceeds as follows.} \hyperref[sec:GPAI_risks]{Section 2} outlines the role of GPAI evaluations in risk assessment, risk mitigation, and AI governance, globally and in the EU. \hyperref[sec:desiderata]{Section 3} calls for GPAI evaluations that follow four desiderata: internal validity, external validity, reproducibility, and portability.   \hyperref[sec:adaptivity]{Section 4} highlights that the mere existence of standards to uphold these desiderata is in itself insufficient and potentially even harmful; instead, standards may best exist within an adaptive framework that can evolve with changes in technology, environment, and risk.  \hyperref[sec:taskforce]{Section 5} proposes the establishment of an EU GPAI Evaluations Standards Taskforce ("the Taskforce") to fulfill the needs highlighted in \hyperref[sec:desiderata]{Section 3} and \hyperref[sec:adaptivity]{Section 4}. \hyperref[sec:provider_commitments]{Section 6} defines GPAI provider commitments to facilitate the success of the Taskforce and highlights that these commitments could be codified in the EU AI Act Codes of Practice. \hyperref[sec:global_impact]{Section 7} discusses the potential global impact of GPAI evaluations standard-setting in the EU. \hyperref[sec:failures]{Section 8} addresses potential challenges and pitfalls for the Taskforce. \hyperref[sec:conclusion]{Section 9} concludes.

\section{GPAI Risks and Evaluations}
\label{sec:GPAI_risks}
\textbf{The development and deployment of general purpose AI (GPAI) models\footnote{Synonymously referred to as foundation models. For the purposes of this paper, we use the term "GPAI models," consistent with the vocabulary of the EU AI Act.} present significant opportunity and considerable risks.} Experts \cite{Hinton2023} and international leaders \cite{AISafetySummit2023} alike have warned of systemic risks of AI systems \cite[Recital 110]{EUAIAct2024}, including:
\begin{itemize}[leftmargin=*]
    \item the misuse \cite{Bommasani2021} of GPAI models, leading to increased scale and severity of cyber attacks,  disinformation, and the development and use of chemical, biological, radiological, and nuclear (CBRN) weapons \cite{Brundage2018};
    \item the malfunctioning or misalignment \cite{Brundage2018} of GPAI models, resulting in loss of control, model autonomy \cite{OpenAI_2023}, and deception; and
    \item broader societal risks related to democratic processes; public and economic security; the dissemination of illegal, false, or discriminatory content; environmental well-being, and inequality \cite{International_Science}.
\end{itemize}

As such, experts have called for the global prioritisation of AI risk mitigation. As the scale and prevalence of AI systems multiply, this task becomes ever more critical.

\textbf{GPAI evaluations play a key role in assessing and mitigating systemic risks \cite{shevlane2023model}.} Current GPAI evaluations \cite{weidinger2023sociotechnical}, or the empirical assessment of the "components, capabilities, behaviour, and impact of an AI system," \cite{weidinger2024holistic} include approaches like benchmarking \cite{bommasani2023holistic}, red-teaming \cite{Ganguli2022b, perez2022red}, and human uplift studies \cite{Mouton_2024}. The results of such evaluations act as a proxy for GPAI model \textit{capability} (how a GPAI model \textit{could} behave in real-world deployment) and \textit{alignment} (how GPAI model \textit{would} behave in real-world deployment) \cite{shevlane2023model}.

\textbf{Accordingly, GPAI evaluations are key in emerging AI governance processes.} Evaluations form the foundation of major GPAI providers' scaling policies and safety frameworks\footnote{Anthropic’s Responsible Scaling Policy, for example, outlines three AI Safety Levels (ASL), whereby model performance on evaluations determines the ASL of a model and the subsequent measures and safeguards to be implemented. DeepMind’s Frontier Safety Framework establishes varying Critical Capability Levels, determined by model performance on model evaluations, at which a range of mitigations shall be triggered. OpenAI’s Preparedness Framework defines model safety scores, as indicated by performance on evaluations, that correspond with key deployment and security decisions.}, including those of Anthropic \cite{Anthropic2023}, DeepMind \cite{Dragan2024}, and OpenAI \cite{OpenAI_2023}. Moreover, in Spring 2024, sixteen industry signatories, including GPAI providers like Amazon, Google, and Meta, committed to "consider[ing] results from internal and external evaluations…set[ting] out thresholds at which severe risks..would be intolerable" and "assess[ing] whether these thresholds have been breached," \cite{AISeoulSummit2024} the latter process in which GPAI evaluations would likely be essential. 

\textbf{From the government side, UK AI Safety Institute (AISI) identifies as one of its three core functions "develop[ing] and conduct[ing] evaluations on advanced AI systems," \cite{AISafetyInstitute2024} and the US AISI aims to "champion the development of empirically grounded tests, benchmarks, and evaluations of AI models, systems, and agents" \cite{USAISI_2024}}.

\textbf{Uniquely, the European Union’s (EU) AI Act mandates that providers must conduct evaluations on GPAI models presenting systemic risk \cite[Article 55]{EUAIAct2024}, making the EU, at the time of writing, the only jurisdiction with the power to mandate and regulate GPAI evaluation practices.} While the AI Office has not yet determined the procedures for GPAI model evaluations \cite{EC_2024}, evaluation results could plausibly trigger regulatory actions, including, by the powers granted to the AI Office, requests for providers to implement safety measures and mitigations \cite[Article 93.1a-b]{EUAIAct2024}, submit additional information \cite[Article 91]{EUAIAct2024}, remove a model from the market (\cite[Article 93.1c]{EUAIAct2024}; or, in extreme cases, the imposition of fines \cite[Article 101]{EUAIAct2024}. As GPAI evaluations hold greater weight in decision- and policy-making – both within and across institutions – establishing and updating standards becomes increasingly important. This task is especially timely in the EU, the world’s first-mover in government-mandated GPAI evaluations.  

\section{Desiderata for Governance-Enhancing GPAI Evaluations}
\label{sec:desiderata}
\textbf{Given their nascent and rapidly evolving state, GPAI evaluations are not currently beholden to standards or best practices} \cite{AISafetyInstitute2024, ApolloResearch2024, pmlr-v235-reuel24a}. To bolster AI governance and mitigate extreme risk within the EU and beyond, standards for GPAI evaluations may seek to uphold a core set of desiderata. This section proposes four such desiderata: internal validity, external validity, reproducibility, and portability. The proposed desiderata draw from evaluations literature and evaluation standards of other sectors\footnote{For example, the European Union’s Registration, Evaluation, Authorisation and Restriction of Chemicals (REACH) outlines regulations for chemical evaluations; and the International Standards Organisation (ISO) and International Electrotechnical Commission (IEC) specify voluntary evaluations standards for medical devices (ISO 13485), cyber and IT products (ISO/IEC 15408), construction materials (ISO 9001), and environmental management systems (ISO 14000).}, with a consideration of the unique characteristics of the GPAI evaluation ecosystem in the EU and beyond. The proposed desiderata, with closed-source GPAI models in mind, inherently foster transparency and can be even more readily achieved with open-source models. These desiderata should not be considered as exhaustive nor definitive, but rather as a starting point for ongoing discussion and refinement. This section defines these desiderata, outlines their policy relevance, surveys the status quo, and provides example standards that might uphold them.  

\subsection{Internal Validity  }
\textbf{Internal validity refers to the extent to which the observed evaluation results represent the truth in the evaluation setting and are not due to methodological shortcomings  \cite{Liao, Mathison, Patino}}. Internal validity is crucial for providing consistent and unbiased insights into the probability and severity of threats presented by GPAI models.  

\textbf{In their current state, evaluations often lack internal validity due to confounding variables, measurement error, and a lack of robust statistical testing.} Benchmark question phrasing \cite{Liang2022, mizrahi_2024} and evaluation structure \cite{Savelka2023, Wang2024}, for example, have been shown to lead to substantially different results, suggesting that such results may be more reflective of methodological artifacts than of the true capabilities of GPAI models. Even holding the same phrasing and structure constant, results of a given evaluation may vary: Lukosiute \cite{Lukosiuteund}, for example, shows that GPQA with chain-of-thought (CoT) prompting yields noisy results across ten runs. These findings underscore the need for well-considered sample sizes and well-powered statistical testing: in the absence of statistical power, meaningful capabilities may go undetected and, conversely, apparently significant capabilities may be exaggerated \cite{Card2024}.  

\textbf{Standards and practices that may promote internal validity include: }
\begin{itemize}[leftmargin=*]
\item Specifications for hypothesis testing, sample sizes \cite{Liao}, random seeds, statistical significance, and statistical power \cite{bowman2021will};
    \item Specification and disclosure of environmental parameters used in evaluations \cite{Liao}; 
    \item Controlling for confounding variables and mitigating spurious correlation (i.e. by randomising question order, ensuring balanced treatment and control groups in experimental settings, etc.); and 
    \item Evidence of the absence of train-test contamination \cite{leakage} and assurance that testing methodologies and test data are not used for GPAI system development.  
\end{itemize}

\subsection{External Validity  }
\textbf{External validity refers to the extent to which evaluation results can be used as a proxy for model behavior in contexts outside of the evaluation environment.} Within external validity lies construct validity, or “how well designed…the experimental setting is in relation to the research claim” \cite{raji2021ai}. External validity is key in bolstering confidence in the use of GPAI evaluations to trigger policy or regulatory decisions, especially in the context of a multi-stakeholder ecosystem \cite{Weidinger_2024}.

\textbf{An extensive body of research documents the ways in which GPAI evaluation results fail to hold under a range of real-world conditions: methods like scaffolding, fine-tuning, and prompting (e.g. single-turn, multi-turn, chain-of-thought) can meaningfully alter GPAI evaluation results \cite{Liang2022, Weber_2023, Bsharat2023, Anthropic2023}, yet no protocols guide the use and documentation of these methods.} Furthermore, GPAI evaluation results often fail to serve as strong proxies for real-world risks. For example, while Meta evaluated Llama 3’s safety by testing its propensity to comply with requests to output harmful content \cite{llama3_modelcard}, its evaluations did not account for the removal of safeguards, which can be cheaply bypassed through fine-tuning \cite{lermen2023lora}. As such, Meta’s evaluation result cannot act as a strong proxy for the real-world risks posed by Llama 3.

\textbf{Standards and practices that may promote external validity include}:  

 \begin{itemize}[leftmargin=*]

\item The clear definition of risks to be measured;\footnote{See, for example, https://airisk.mit.edu/.}
    \item Input of domain expertise in the design of evaluations, ensuring that the evaluation is a well-considered proxy for the broader real-world task or risk \cite{Liao, Tabassi_2023};
    \item The use of strong elicitation in GPAI evaluations, including fine-tuning, scaffolding, white-box adversarial attacks \cite{casper2024black}, and prompt engineering; and
    \item Documentation of factors informative of the evaluation’s external validity across contexts \cite{glasgow2006external}, including:  
    \begin{itemize}
        \item The environmental conditions in which the evaluation is run and the ways in which it converges with and diverges from the real-world context \cite{Liao}; and  
        \item In the case of red-teaming and adversarial testing, details on the sourcing and vetting of domain experts to partake in the evaluation and the level and duration of access to GPAI models provided to these experts for the purpose of the evaluation.  
    \end{itemize}
\end{itemize}

\subsection{Reproducibility   }
\textbf{Reproducibility refers to the ability to obtain consistent evaluation results using the same input data, computational methods, code, and evaluation conditions \cite{NAP25303}.} Reproducibility is key in bolstering confidence in the use of GPAI evaluations to trigger policy or regulatory decisions, especially in the context of a multi-stakeholder ecosystem \cite{weidinger2023sociotechnical}. Little attention has been given to date to the reproducibility of GPAI evaluations, though a need exists \cite{Ganguli2023}. While reproducibility may not be applicable to all types of evaluations, its use should be strongly considered and its absence should be carefully justified.  

\textbf{Standards and practices that may promote reproducibility include: }
 \begin{itemize}[leftmargin=*]
\item Definition of environments and evaluations for which reproducibility can enhance risk assessment, risk mitigation, and AI governance; 
\item For the above-indicated environments and evaluations: 
    \begin{itemize}
        \item secure release of evaluation data, including input data, random seed, and output data for steps that are nondeterministic and cannot be reproduced \cite{NAP25303}
        \item secure release of evaluation code \cite{Goodman2016}; and
        \item documentation of evaluation methodology, evaluation environment, computation al environment (i.e. hardware architecture, operating systemic, and library dependencies), elicitation methods, statistical testing, and analysis \cite{Nosek_2018, NAP25303}.
    \end{itemize}
\end{itemize}

\subsection{Portability}
\textbf{Portability refers to the ability of a range of stakeholders to consistently and seamlessly implement and assess GPAI evaluations across distinct institutions and hardware environments through, for example, accessible evaluation software.} Portability facilitates both collaboration and external scrutiny, the latter of which can avail more reliable information by verifying GPAI provider claims and revealing new insights \cite{Anderljung2023, Manheim_2024}. Portability may furthermore enhance resource efficiency for governments and third parties, foster knowledge spillovers among stakeholders, and reduce pressure for industry concentration from regulation costs \cite{interoperability}. 

\textbf{The EU AI Act \cite{EUAIAct2024}) and the Frontier AI Safety Commitments \cite{AISeoulSummit2024} highlight the advent of multi-stakeholder evaluation and, by extension, the need for portability.} The EU AI Act outlines at least three distinct parties that may run GPAI evaluations: GPAI providers \cite[Article 55.1a]{EUAIAct2024}, the EU AI Office \cite[Article 92.1]{EUAIAct2024}, and qualified independent experts on behalf of the AI Office \cite[Article 92.2]{EUAIAct2024}. Separately, signatories of the Frontier AI Safety Commitments commit to "consider[ing] results from internal and external evaluations as appropriate, such as by independent third-party evaluators, their home government, and other bodies their governments deem appropriate" \cite{AISeoulSummit2024}. Few common frameworks to date facilitate portability of evaluations. METR’s Task Standard \cite{METR_taskstandard} standardizes task-based evaluations to facilitate efficiency, METR’s Vivaria \cite{METR_vivaria} provides a platform for running evaluations and conducting elicitation research, and the UK AI Safety Institute’s Inspect \cite{INSPECT} allows users to assess and score GPAI model capabilities in a range of areas, While these platforms offer a starting point for portability, substantial scope remains to promote more comprehensive and well-defined standards across evaluation types and contexts \cite{reuel2024open}. 

\textbf{Standards and practices that may promote portability include:} 
 \begin{itemize}[leftmargin=*]
\item The development and use of software that: 
\begin{itemize}
    \item makes evaluations virtually "plug and play," reducing the engineering effort needed to implement another actor's evaluations;  
    \item is model- and agent-architecture-agnostic;\footnote{e.g. evaluations that do not depend on a particular model or agent architecture, and can therefore be re-used with different models and agent architectures.  }
    \item facilitates swapping out and combining particular post-training enhancements between different evaluators; and  
    \item to address privacy constraints, where relevant, facilitates evaluation implementation in a privacy-preserving manner (e.g. without leaking additional information about a model or an evaluation to the other party) \cite{bluemke}.
\end{itemize}
 \end{itemize}

\section{The Need for Adaptive Standards}
\label{sec:adaptivity}
\textbf{Establishing standards for GPAI evaluations can promote desiderata like internal validity, external validity, reproducibility, and portability \cite{weidinger2024holistic}.} However, if overly rigid and static, standards may threaten the innovation and progress of GPAI evaluations. Given that GPAI evaluations are a nascent and rapidly evolving field in which best practices are subject to rapid technological and environmental advancements, the existence of standards in itself is insufficient and potentially even harmful if not paired with adaptivity \cite{grote_2015}. 

\textbf{Adaptivity refers to the ability of evaluation standards to anticipate and adjust nimbly to changes in technology, environment, and risk.} GPAI evaluations are perpetually subject to change as landscapes shift, human interaction with models evolve, models gain new capabilities with potentially abrupt performance jumps, and new harm vectors emerge \cite{Ganguli2022a}. Furthermore, evaluations like benchmarks may become less reliable over time due to increasing likelihood of leakage into the training dataset \cite{weidinger2024holistic}. Hence, in order to most effectively contribute to risk mitigation, GPAI evaluations methodology and measurement must continuously evolve. 

\section{GPAI Evaluations Standard Taskforce}
\label{sec:taskforce}
\textbf{To address the need for adaptive standards, we propose a GPAI Evaluations Standards Taskforce ("the Taskforce"), to be housed within institutions established by the EU AI Act.} The Taskforce is a vetted body of independent researchers to maintain and update GPAI evaluation standards. Given the potential regulatory leverage of GPAI evaluations in the EU and the evolving nature of GPAI evaluations, a dedicated Taskforce can promote evaluation standards that foster desiderata like internal validity, external validity, reproducibility, and portability in an evolving manner in the EU and potentially beyond. This proposal follows a broader call for Audit Standards Boards \cite{Manheim_2024} and proactive risk management in dynamic fields \cite{Rasmussen_2000}.

\subsection{Institutional Setting}
\textbf{The provisions of the EU AI Act provide the scope for the Taskforce to exist.} This proposed Taskforce placement within the EU builds from \cite{weidinger2024holistic}, which suggests that "institutionalized scientific panels such as...the EU’s newly established AI Office...could serve the need to continually update...best practices for model evaluations in light of new technological and scientific advances." Within the provisions of the EU AI Act, the Taskforce could sit within two potential institutional settings, outlined below:  

\tcbset{
  boxstyle1/.style={
    enhanced,
    colback=purple!10,
    colframe=purple!70!black,
    coltitle=white,
    fonttitle=\bfseries,
    title style={fill=purple!70!black},
    arc=0mm,
    outer arc=0mm,
  },
  boxstyle2/.style={
    enhanced,
    colback=blue!10,
    colframe=blue!70!black,
    coltitle=white,
    fonttitle=\bfseries,
    title style={fill=blue!70!black},
    arc=0mm,
    outer arc=0mm,
  }
}

\begin{tcolorbox}[boxstyle1, title=IS1: The Scientific Panel of Independent Experts]
\textbf{The Scientific Panel of Independent Experts ("the Panel")}, established by \cite[Article 68 of the EU AI Act]{EUAIAct2024}, is tasked with a range of responsibilities, including the "development of tools and methodologies for evaluating capabilities of general-purpose AI models and systems." \hyperref[sec:appendixA]{Appendix A} outlines the responsibilities of the Panel, notes Taskforce-relevant tasks, and highlights Panel duties that would be symbiotic with the proposed Taskforce work.
\end{tcolorbox}

\begin{tcolorbox}[boxstyle2, title=IS2: The Advisory Forum]
\textbf{The Advisory Forum ("the Forum"),} established by \cite[Article 67 of the EU AI Act]{EUAIAct2024}, is tasked with providing technical expertise and advising the European Artificial Intelligence Board and the Commission. The Forum "may establish standing or temporary sub-groups as appropriate for the purpose of examining specific questions related to the objectives of this Regulation." By mandate, the Forum includes permanent advisory members from the European Committee for Standardisation (CEN), which holds a technical cooperation agreement with the International Standards Organisation (ISO) via the Vienna Agreement. This framework paves the path for potential dialogue between GPAI evaluations standards at the Taskforce-, EU-, and international-level.
\end{tcolorbox}

\textbf{Relative to alternative placements,\footnote{Alternatives include, for example, the establishment of GPAI evaluations standards by third-party evaluators directly, GPAI providers directly, CEN/CENELEC, and the International Standards Organisation (ISO).} IS1 and IS2 offer conditions particularly conducive to developing and implementing GPAI evaluation standards,} including timely access to existing institutional infrastructure; the ability to convene expert pools that represent diverse, multi-stakeholder perspectives; regulatory ties to GPAI provider and models; the legal basis for implementation of GPAI evaluations and standards; and strong institutional ties to international-standard-setting organisations.  

\textbf{Precedents for such a Taskforce within the EU exist} and include that established by Commitment 27 of the 2022 Strengthened Code of Practice on Disinformation \cite{EC_2022} in which signatories including Google, Meta, and Microsoft commit to "provide vetted researchers with access to data necessary to undertake research on Disinformation by developing, funding, and cooperating with an independent, third-party body that can vet researchers and research proposals."

\subsection{Structure and Duties}
\textbf{The Taskforce could be composed of technical experts from academia, civil society, third-party model audit providers, regulators, and experts from government serving in a personal capacity}, with Taskforce members  furthermore fulfilling the criteria outlined in \cite[Article 68.2]{EUAIAct2024}, if falling within IS1, or \cite[Article 67.2]{EUAIAct2024}, if falling within IS2. The latter additionally allows for the involvement of GPAI industry experts. The Taskforce could facilitate regular pathways for input by a range of non-Taskforce members, including GPAI industry experts.\footnote{ If falling within IS1, by way of \cite[Article 68.2]{EUAIAct2024}, Taskforce members must be independent from industry. Involving industry insights via non-Taskforce input is valuable.}

\textbf{Key Taskforce duties, drawing from the framework of the EU AI Act and building towards the desiderata highlighted in \hyperref[sec:desiderata]{Section 3}, could include:}

\subsubsection{Harmonisation of Risk Taxonomies and GPAI Model Evaluation Methodologies}
\begin{itemize}[leftmargin=*]
    \item Closely collaborating with the Scientific Panel of Independent Experts to promote a regularly updated comprehensive taxonomy and enumeration of GPAI systemic risks ("the Risk Taxonomy"), to be initially established by the Codes of Practice \cite[Recital 116]{EUAIAct2024}; 
    \item Outlining and updating concrete examples of and standards for qualitative and quantitative experiments that would demonstrate the presence of the risk;
    \item Ensuring the continuous updating of and harmonisation between the Risk Taxonomy and GPAI evaluations standards; and
    \item Ensuring that updates to the taxonomy outline precise descriptions of risks, taking into consideration advancements in technology and the evolution of external environments, and drawing from resources including but not limited to provider safety cases reports, scaling risk management policies, capability forecast reports, and incident reporting documentation.
\end{itemize}

\subsubsection{Standard-setting for GPAI Evaluations}
\begin{itemize}[leftmargin=*]
    \item Regularly\footnote{Such as every twelve months or by qualified alert.} publishing updated standards for GPAI model evaluations to guide evaluations implemented by providers \cite[Article 55.1.a]{EUAIAct2024} and by the AI Office and third-party evaluators \cite[Article 92]{EUAIAct2024} towards internal validity, external validity, reproducibility, and portability; and
    \item Defining environments and evaluations for which standards are not safety-enhancing.
\end{itemize}

\subsubsection{Quality-control of Evaluations}
\begin{itemize}[leftmargin=*]
    \item Setting standards for the third-party evaluator conduct, following the proposal in \cite{raji2022outsider} for an audit oversight board modelled after the Public Company Accounting Oversight Board (PCAOB);
    \item Vetting, approving, and periodically\footnote{Such as every twelve months or by qualified alert} reviewing the fitness of third-party evaluators; and
    \item Auditing the methodologies and results of GPAI evaluations implemented by providers, the AI Office, and third-party evaluators both pre- and post-deployment.
\end{itemize}

Finally, while not a primary duty, the Taskforce should facilitate the harmonisation of GPAI model evaluation standards across GPAI providers and international bodies, including but not limited to standards organisations and AI Safety Institutes, where possible, provided such efforts do not compromise the independence of the Taskforce or the quality of its standards.

\section{GPAI Provider Commitments}
\label{sec:provider_commitments}
\textbf{The success of the Taskforce hinges upon its direct exchange with GPAI provider experts to support access and provide relevant information} \cite{weidinger2024holistic, Anderljung2023}. The Frontier AI Safety Commitments \cite{AISeoulSummit2024} lay the groundwork for this collaboration\footnote{Signatories commit to “work toward information sharing,” “incentivize third-party discovery and reporting of issues and vulnerabilities,” “publicly report model or system capabilities, limitations, and domains of appropriate and inappropriate use;” and “prioritise research on societal risks posed by frontier AI models and systems.”} and the EU AI Act Code of Practice \cite{EC_2024}, to be finalized by May 2025, could formally codify GPAI provider commitments to the Taskforce. To concretely support the development and ongoing success of the Taskforce, GPAI providers can commit to working with the relevant organisations (e.g. the European Commission, the EU AI Office, Civil Society, Data Protection Authorities) to establish and sustain the Taskforce. In particular, this requires commitments to the provision of documentation and model and data access. Proposed provider commitments are listed in \hyperref[sec:appendixB]{Appendix B}.

\section{Potential for Global Impact}
\label{sec:global_impact}
\textbf{The EU could plausibly pave the pathway for coordinated international standards of GPAI evaluations via a \textit{de jure} Brussels effect (the adoption of EU standards by foreign governments) and a \textit{de facto} Brussels effect (the unilateral regulation of GPAI provider practices) \cite{Siegmann_2022, bradford2020brussels}.}
\break
\break
\textbf{\textit{De jure}, GPAI evaluations standards in the EU could inform norms and standards internationally by way of collaboration with AI Safety Institutes and/or via the CEN-ISO pipeline, as discussed in IS2  (\hyperref[sec:taskforce]{Section 5})}. 
Evaluations may play a key policy role beyond the EU, including in the UK and US \cite{AISafetyInstitute2024, AISeoulSummit2024}. As the potential first-mover in GPAI evaluations standards – and the only current jurisdiction able to mandate and regulate GPAI evaluation practices – the EU could plausibly pave the pathway for coordinated international standards. As AI Safety Institutes expand and synergise, the alignment of evaluation standards and the exchange of results could facilitate efficiency, returns to scale, and specialization \cite{variengien2024ai}.
\break
\break
\textbf{\textit{De facto}, GPAI evaluations standards could shape GPAI provider practices internationally.} A priori, anticipation and awareness of standards for GPAI evaluations mandated by the EU could preemptively shift GPAI provider behavior. Post-hoc, standards could impact evaluation results and trigger regulatory decisions. Both cases may, in turn, prompt changes in GPAI model development and deployment in the EU and potentially beyond. Previous estimates projected the EU’s AI market share to be between 15\% \cite{IMF2022} - 22\% \cite{EuropeanCommission2021}; depending on regulatory frameworks, this market share could provide strong incentives for GPAI providers to develop and deploy GPAI models that are provably safe, as measured by GPAI evaluations and accompanying standards. 

\section{Potential Causes of Failure and Additional Approaches}
\label{sec:failures}
\textbf{While the Taskforce could improve AI risk assessment and mitigation in the best case scenario, it could also face challenges and downfalls.} Beyond standard issues of funding, regulatory capture \cite{dal2006capture, ramanna2015political, Wu2023capture}, and misaligned incentives \cite{guha2023ai}, this section draws from literature to highlight potential causes of failure for the Taskforce and propose related approaches:

\begin{itemize}[leftmargin=*]
    \item \textbf{Stifled acquisition of talent:} the Taskforce may face challenges in recruiting sufficiently talented experts. This could be exacerbated by an inability to match industry salaries or quality of life, restrictions on hiring industry-affiliated researchers, or other hiring process restrictions resulting in long hiring timelines or an inability to flexibly shape positions. Ensuring sufficient salary, nimble workstyle, and hiring flexibility could address these challenges  \cite{pmlr-v235-reuel24a}. 
    
    \item \textbf{Heavy bureaucratic processes:} while adaptivity is highlighted as a priority, in practice the Taskforce could struggle to keep pace with rapidly advancing AI capabilities. Bureaucratic processes, transparency, conflict of interest requirements, and the need for consensus could slow down the general pace of work and restrict the ability of the Taskforce to gather information from experts  \cite{kauffman}. Granting the Taskforce the authority and political capital to establish its own processes from first principles instead of inherent processes from established industry players or government could address this. 
\end{itemize}

In addition to the Taskforce merely failing to achieve its mission, two particularly prominent ways in which its efforts could backfire include:

\begin{itemize}[leftmargin=*]
    \item \textbf{Increased friction for evaluators:} standards will almost certainly create some extra processes for evaluators within GPAI providers as well as third parties. If this friction is too great, it could disincentivise new evaluators from entering the field at a time when talent is a critical bottleneck, and it could reduce the ability of evaluators to experiment with new methodologies and advance this critical risk management science \cite{brunsson}.
    
    \item \textbf{False sense of security:} the existence of standards could create a false sense that AI risks are being adequately managed, potentially reducing vigilance or investment in other areas of risk management that might be more promising \cite{mukobi_doubt, power}. This is especially true insofar as it will be difficult to assess the effectiveness of these standards.
\end{itemize}

In light of the potential failures or downsides of the Taskforce, it is prudent to also consider other approaches to AI risk management, both as complements to and substitutes for the Taskforce, including but not limited to:

\begin{itemize}[leftmargin=*]
    \item \textbf{Outcome-based regulation:} focus on regulating the outcomes and impacts of AI systems rather than specific evaluation methodologies \cite{coglianese}.
    \item \textbf{Mandatory information sharing:} require AI companies to share more detailed information about their systems and internal evaluation processes, fostering transparency, independent of specific standards \cite{Ganguli2023}.
    \item \textbf{Focus on infrastructure:} invest in shared testing infrastructure that can be used by multiple stakeholders to evaluate AI systems.
    \item \textbf{Bottom-up standards development:} encourage the development of standards through a distributed, community-driven process \cite{morris} involving a wide range of stakeholders, such as regularly posing open challenges or competitions to identify novel risks or evaluation methods, supplementing the work of a permanent Taskforce.
    \item \textbf{AI bounty programs:} establish programs that reward individuals or organisations for identifying potential risks or vulnerabilities in AI systems \cite{heikkila_2022}.
\end{itemize}

\section{Conclusion}
\label{sec:conclusion}
\textbf{While GPAI evaluations have been proposed as a central tool in assessing and mitigating systemic risks posed by GPAI development and deployment, no established standards exist to date to promote their quality and legitimacy.} This paper proposes an EU GPAI Evaluations Standards Taskforce to develop and adapt standards for desiderata of GPAI evaluations, including internal validity, external validity, reproducibility, and portability. Amidst the rapidly evolving state of GPAI evaluations and constant shifts in environments, technology, and risks, a Taskforce could promote relevant and governance-enhancing standards for effective risk assessment and mitigation in the EU and beyond. The success of the Taskforce and its responsibilities, as outlined in \hyperref[sec:taskforce]{Section 5}, rely on collaboration with GPAI providers, as outlined in \hyperref[sec:provider_commitments]{Section 6}. The impact of the Taskforce, for better or for worse, should not be underestimated: by way of \textit{de facto} and \textit{de jure} Brussels effects, the Taskforce could influence GPAI evaluations standards and GPAI safety at an international scale. As such, any design and management of the Taskforce should heed the potential pitfalls and recommendations highlighted in \hyperref[sec:failures]{Section 8}.
\break
\break
\textbf{This proposal for an EU GPAI Evaluation Standards Taskforce rests on four key assumptions:} 1) well-formulated GPAI evaluations standards can support effective risk assessment, risk mitigation, and AI governance, 2) developing standards within institutions and environments capable of mandating and implementing standards increases the legitimacy and impact of those standards, 3) when establishing standards, the interactions fostered by a dedicated, multi-stakeholder Taskforce can ultimately increase the quality of standards, relative to many distinct, uncoordinated bi-lateral and multilateral interactions, and 4) the benefits to AI governance and public safety of establishing a GPAI Evaluations Standards Taskforce outweigh the costs.  
\break
\break
\textbf{This paper draws primarily on literature, current AI policy, and the authors' expertise. It faces the limitations endemic to the nascent field of GPAI evaluations and AI policy more broadly:} sparse evidence, rapid advancements, and, accordingly, a lack of conclusive research from which to draw. Future research may employ qualitative methods like interviews to draw insights and recommendations from the multi-stakeholder GPAI evaluations community; as well as quantitative methods like surveys, impact evaluations, and forecasting to more rigorously estimate the impacts of an EU GPAI Evaluations Standards Taskforce on the state of risk assessment, risk mitigation, and AI governance in the EU and beyond.
\clearpage
\section*{Impact Statement}

While GPAI evaluations are crucial for assessing and mitigating systemic risks, no standards currently exist to ensure their quality and legitimacy. This paper proposes an EU GPAI Evaluations Standards Taskforce to develop and adapt standards for robust, reproducible, and interoperable evaluations. The Taskforce, to be housed within institutions established by the EU AI Act, could enhance AI governance by providing adaptive frameworks for evaluations, on which policy-making processes increasingly rely. However, the success of the Taskforce may be stifled by the rapid pace of AI progress, potential bureaucratic obstacles, and the potential inadvertent stifling of safety-related innovation due to rigid standards. Despite these hurdles, establishing such a Taskforce could  contribute to the responsible development and deployment of GPAI systems, both within and beyond the EU, and ultimately facilitate more effective governance in this critical domain.

\section*{Appendix A: Taskforce Integration in the Panel}
\label{sec:appendixA}
Article 68.3 \cite{EUAIAct2024}, below, outlines the tasks of the Scientific Panel of Independent Experts. \textbf{T} demarcates tasks that could be partially or fully delegated to the Taskforce, given the tasks' direct relevance to GPAI evaluations standards. \textbf{TP} demarcates tasks that hold complementarities with GPAI evaluations but are not exhaustively fulfilled by the duties outlined in \hyperref[sec:taskforce]{Section 5}. Thus, the Taskforce could maintain a close dialogue with the Panel on these tasks, particularly on Task a.i., given its role in robustness and construct validity in GPAI evaluations. \textbf{P} demarcates tasks outside of the Taskforce's mandate.

\begin{mdframed}
(a) supporting the implementation and enforcement of this Regulation as regards general-purpose AI models and systems, in particular by:

\hspace{1em}(i) alerting the AI Office of possible systemic risks at Union level of GPAI models, in accordance with \cite[Article 90]{EUAIAct2024} (\textbf{TP});

\hspace{1em}(ii) contributing to the development of tools and methodologies for evaluating capabilities of GPAI models and systems, including through benchmarks (\textbf{T});

\hspace{1em}(iii) providing advice on the classification of GPAI models with systemic risk (\textbf{P});

\hspace{1em}(iv) providing advice on the classification of various GPAI models and systems (\textbf{P});

\hspace{1em}(v) contributing to the development of tools and templates (\textbf{TP});

(b) supporting the work of market surveillance authorities, at their request (\textbf{P});

(c) supporting cross-border market surveillance activities as referred to in \cite[Article 74.11]{EUAIAct2024}, without prejudice to the powers of market surveillance authorities (\textbf{P});

(d) supporting the AI Office in carrying out its duties in the context of the Union safeguard procedure pursuant to \cite[Article 81]{EUAIAct2024} (\textbf{P}).
\end{mdframed}
\section*{Appendix B: Provider Commitments}
\label{sec:appendixB}
\subsection*{Documentation}
A comprehensive understanding of evaluations results relies on intimate knowledge of the evaluations process \cite{Ganguli2023}\footnote{Consider, for example, specialised evaluations like Bias Benchmark for QA (BBQ), which measure social biases and require nuanced interpretation. Human evaluations, while valuable for assessing real-world performance, are subject to variability based on evaluator characteristics and the inherent subjectivity of human judgment. Model-generated evaluations, while efficient, may inherit biases or inaccuracies from the models that created them.}. Access to detailed documentation is crucial to account for nuances and potential pitfalls, interpret evaluations, and develop standards accordingly. 

GPAI providers commit to draw-up and keep up-to-date sufficient documentation on their evaluation results and any additional information needed to rigorously audit these results. Sufficient documentation is understood to include, upon the determination of the AI office and proportional to the risk being assessed, the required methods and the security precautions taken by the Taskforce:  
\begin{itemize}[leftmargin=*]
    \item Description of models and methods as outlined in \cite[Annex XI.1]{EUAIAct2024}, including but not limited to training data, training compute used, and number of perameters;\footnote{This information often includes valuable trade secrets, and providing it to outsiders risks leaking it to outsiders. On the other hand, knowing these details can be crucial for verifying safety cases and evaluation results. The risk associated with sharing this information should be weighed against the improved oversight that it enables.}
    \item Documentation outlined in Annex XI.2.2.1 \cite{EUAIAct2024}, including specifically:
    \begin{itemize}
        \item Scores on individual benchmarks and other developers' evaluations;
        \item Description of evaluations and benchmarks;
        \item Model responses on benchmarks; and
        \item Descriptions and examples of the data included in the benchmarks.
    \end{itemize}
    \item Documentation outlined in Annex XI.2.2 \cite{EUAIAct2024}, including specifically:
    \begin{itemize}
        \item Description of capability elicitation procedures used in the benchmarks; 
        \item Calendar-time and employee-hours spent conducting the evaluation; and
        \item The procedure used to fine-tune the model.
    \end{itemize}
    \item Documentation outlined in Annex XI.2.3 \cite{EUAIAct2024}, including specifically:
    \begin{itemize}
        \item The scaffolding and tools available to the model during inference; and
        \item Details about the model's system architecture.
    \end{itemize}
\end{itemize}

GPAI providers additionally commit to providing updated documentation to the Taskforce, including but not limited to safety cases, scaling risk management policies, capability forecast reports, and incident reporting documentation, as outlined in Article 55.1.c and Recital 115.
GPAI providers additionally commit to providing updated documentation to the Taskforce, including but not limited to safety cases, scaling risk management policies, capability forecast reports, and incident reporting documentation, as outlined in Article 55.1.c and Recital 115 \cite{EUAIAct2024}.

\subsection*{Model and data access}
Comprehensively implementing, drawing insights from, and setting standards for GPAI evaluations require sufficient model access. Sufficient access is understood to include, upon the determination of the Taskforce and proportional to the risk being assessed, the required methods and the security precautions taken by the Taskforce:

\begin{itemize}[leftmargin=*]
    \item The ability to use the GPAI system in the way that it will be made available to customers (e.g. prompting it for text or image responses);
    \item White box model access \cite{casper2024black}; 
    \item Access to versions of the GPAI system that lack technical safety mitigations, as these can prevent evaluators from exploring the full range of an AI system's capabilities \cite{UK_DSIT}
    \item Through an application programmer interface (API), the ability to fine-tune a GPAI system \cite{trask2020beyond, shevlane2022structured, bucknall2023structured}; 
    \item Access to other models in the model family \cite{bucknall2023structured}; 
    \item Access to individual components of the GPAI system, including the core model (via API) and other software components, such as moderation filters, system prompts, and available plug-ins allowing additional capabilities like web browsing and code execution \cite{bucknall2023structured}; and
    \item The data used to train the GPAI model, and/or meta-data, descriptions, and examples of the training data.
\end{itemize}

Some forms of enhanced access may increase the risk that GPAI providers’ intellectual property and trade secrets are leaked. To fulfil Article 78  of the EU AI Act \cite{EUAIAct2024}, providers commit to working with the  EU AI Office to provide sufficient access while minimising information security concerns. Techniques for achieving this objective may include:

\begin{itemize}
    \item Structured API access, providing "de facto white-box" capabilities to third parties without giving direct access to model parameters \cite{shevlane2022structured}; 
    \item Physical solutions involving secure research environments on a GPAI provider's campus, allowing unrestricted white-box access while minimising risks of leaks;
    \item Legal mechanisms that draw on practices from other industries with audits can be used to enforce confidentiality and hold third parties accountable.
\end{itemize}

\bibliography{ref}
\bibliographystyle{plainnat}

\end{document}